\documentclass[english,twocolumn]{revtex4-1}
\usepackage{amsmath}
\usepackage{mathptmx}
\usepackage{newtxmath}

\usepackage[T1]{fontenc}
\usepackage[latin9]{inputenc}
\usepackage{stmaryrd}
\usepackage{graphicx}
\usepackage{wasysym}

\makeatletter
\newcommand{\expv}[1]{\left\langle #1\right\rangle}

\makeatother

\usepackage{babel}
\begin{document}

\title{Cover time for random walks on arbitrary complex networks}

\author{Benjamin F. Maier}
\email{bfmaier@physik.hu-berlin.de}

\affiliation{Robert Koch-Institute, Nordufer 20, D-13353 Berlin}

\affiliation{Department of Physics, Humboldt-University of Berlin, Newtonstraße
15, D-12489 Berlin}

\author{Dirk Brockmann}

\affiliation{Robert Koch-Institute, Nordufer 20, D-13353 Berlin}

\affiliation{Institute for Theoretical Biology, Humboldt-University of Berlin,
Philippstr. 13, D-10115 Berlin}
\begin{abstract}
We present an analytical method for computing the mean cover time
of a discrete time random walk process on arbitrary, complex networks.
The cover time is defined as the time a random walker requires to
visit every node in the network at least once. This quantity is particularly
important for random search processes and target localization on network
structures. Based on the global mean first passage time of target
nodes we derive a method for computing the cumulative distribution
function of the cover time based on first passage time statistics.
Our method is viable for networks on which random walks equilibrate
quickly. We show that it can be applied successfully to various model
and real-world networks. Our results reveal an intimate link between
first passage and cover time statistics and offer a computationally
efficient way for estimating cover times in network related applications.
\end{abstract}
\maketitle

\section{Introduction}

Random walks have been studied extensively for more than a century
and emerged as an efficient descriptive model for spreading and diffusion
processes in physics, biology, social sciences, epidemiology, and
computer science \cite{oksendal_stochastic_1992,berg_random_1993,barrat_dynamical_2008,Brockmann:2008hb,newman_networks:_2010,klafter_first_2011,masuda_random_2016}.
Because of their wide applicability and relevance to dynamic phenomena,
random walk processes have become a topic of interest particularly
for analyses of dynamics on complex networks \cite{masuda_random_2016}.
The calculation of a resistor network's total resistance \cite{newman_networks:_2010},
synchronization phenomena in networks of coupled oscillators \cite{barrat_dynamical_2008},
the global spread of infectious diseases on the global air traffic
network \cite{Hufnagel:2004kta,Brockmann:2013ud,iannelli_effective_2017}
and ranking the importance of single websites in the world wide web
\cite{page_pagerank_1999} are just a few examples of systems that
have been investigated based on concepts derived from random walk
theory. 

Especially important is the understanding of temporal aspects of stochastic
processes and how different network structures influence the equilibration
process. Consequently, a lot of theoretical work focused on understanding
the connection between network structure and relaxation time scales
or first passage times (FPTs), the time it takes a single walker to
travel from one node to another. Both, relaxation and first passage
times quantify different aspects but fail to capture the characteristic
time a walker requires on average to visit \textit{every} node in
a network, which is captured by the \textit{cover time} of the process.
This quantity, however, has important practical applications from
biology to computer science, for instance, for estimating how long
it will take to distribute a chemical or a certain commodity to every
node in a network or as a measure for navigability in multilayer transportation
networks \cite{domenico_navigability_2014}. Researchers have been
able to derive analytically asymptotic results for the cover time
for some model networks, e.g.~complete graphs, Erd\H{o}s\textendash Rényi
(ER) and Barabási\textendash Albert (BA) networks \cite{cooper_cover_2007,cooper_cover_2007-1,lovasz_random_1996}.
Yet, only few analytical or heuristic results concerning the mean
cover time of real-world networks have been established, it is thus
unclear how a real-world network's mean cover time is related to other
temporal features of random walks on these networks and how their
structure and topological features may impact cover time statistics.

In the following, we present a theoretical approach that predicts
the cover time on arbitrary complex networks using only FPT statistics.
We show that for networks on which random walks equilibrate quickly
(specified below), the cover time can be estimated accurately by the
maximum of a set of FPTs drawn from the ensemble of FPT distributions
of all target nodes. Our method's predictions are in excellent agreement
with results provided by computer simulations for a variety of real-world
networks, as well as for ER networks, BA networks, complete graphs
and random $k$-regular networks. We also show that our method fails
when the conditions of rapid relaxation are violated, e.g. for networks
embedded in a low-dimensional space with short-range connection probability.

\section{Theory}

\subsection{Random walks and first passage times}

\label{subsec:simple_random_walk}The foundation of our analysis is
an unweighted, undirected, network composed of $N$ nodes, $E$ links
and adjacency matrix $A_{vu}$ with $A_{vu}=1$ if node $u$ and $v$
are connected and $A_{vu}=0$ if not. On this network, we consider
a simple discrete time random walk that starts on an initial node
$u$ at time $t=0$. At every time step, the walker jumps randomly
to an adjacent node $v$. The process is repeated indefinitely and
is goverened by the master equation
\[
P_{v}(t+1)=\sum_{u=1}^{N}W_{vu}P_{u}(t),
\]
where $P_{u}(t)$ is the probability that the walker is at node $u$
at time $t$, $W_{vu}=A_{vu}/k_{u}$ is the transition probability
of the walker going from node $u$ to node $v$ in one time step,
and $k_{u}=\sum_{v=1}^{N}A_{vu}$ is the degree of node $u$. We assume
that the network has a single component, so every node can be reached,
in principle, from every other node. Generally, this process will
approach the equilibrium $P_{v}^{\star}=k_{v}/2E.$ 

Central questions for random walks are often connected to first passage
times (FPT), e.g. the mean first passage time $\tau_{vu}$ (MFPT)
between two nodes $u$ and $v$. This time is defined as the mean
number of steps it takes a random walker starting at node $u$ to
first arrive at node $v$. Another important quantity is the global
mean first passage time of node $v$ (GMFPT), obtained by averaging
the MFPT over all possible starting nodes:
\begin{equation}
\tau_{v}=\frac{1}{N-1}\sum_{u\neq v}\tau_{vu}.\label{eq:GMFPT}
\end{equation}
The GMFPT can be used as a measure of centrality for node $v$ since
a node that is quickly reachable from anywhere may be interpreted
to be ``important''. Passage times have been well analyzed and can
be computed efficiently from network properties. Given the unnormalized
graph Laplacian 
\[
L_{vu}=k_{v}\delta_{vu}-A_{vu},
\]
where $\delta_{vu}$ denotes Kronecker's delta, the MFPT between two
nodes can be computed by spectral decomposition~\cite{newman_networks:_2010,lin_mean_2012}.
Given the operator's eigenvalues $0=\lambda_{1}<\lambda_{2}\leq...\leq\lambda_{N}$
and corresponding orthonormal eigenvectors $\mu_{i}=(\mu_{i1},\mu_{i2},...,\mu_{iN})^{T}$,
one can compute node $v$'s exact GMFPT as
\begin{equation}
\tau_{v}^{\mathrm{ex}}=\frac{N}{N-1}\sum_{i=2}^{N}\frac{1}{\lambda_{i}}\left(2E\mu_{iv}^{2}-\mu_{iv}\sum_{n}k_{n}\mu_{in}\right).\label{eq:GMFPT_Laplacian}
\end{equation}
A computationally more efficient method estimates the GMFPT by its
lower bound, which is given by
\begin{equation}
\tau_{v}^{\mathrm{es}}\geq\frac{N\expv k}{k_{v}}\frac{1}{1-\expv k^{-1}},\label{eq:GMFPT_lower_bound}
\end{equation}
if the process equilibrates quickly, i.e.~when the relaxation time
fullfills $t_{\mathrm{rlx}}\ll N$, which holds within small relative
errors for Erd\H{o}s\textendash Rényi (ER) and Barabási\textendash Albert
(BA) networks, as well as for a variety of real-world networks \cite{lau_asymptotic_2010}.
The relaxation time of a random walk on a network can be bounded from
below using the second smallest eigenvalue $\lambda_{2}$ of $L_{vu}$
\cite{mohar_applications_1997} as 
\begin{equation}
t_{\mathrm{rlx}}\geq\lambda_{2}^{-1}.\label{eq:t_rlx}
\end{equation}
Another temporal characteristic with practical relevance is the mean
cover time $T_{u}$, defined as the mean number of steps it takes
a random walker starting at node $u$ to visit every other node at
least once. For various network models simple heuristics concerning
the asymptotic scaling of the mean cover time as a function of network
size $N$ have been derived \cite{cooper_cover_2007,cooper_cover_2007-1,lovasz_random_1996}.
For ER, BA and fully connected networks it was shown that 
\[
\left\langle T\right\rangle \sim\alpha\,N\log N,
\]
with network specific prefactor $\alpha$. Here, $A_{N}\sim B_{N}$
means $\lim_{N\rightarrow\infty}A_{N}/B_{N}=1$ as in Ref.~\cite{cooper_cover_2007-1}.
Such scaling relationships are useful for comparative analyses, e.g.~when
networks for different sizes of the same class are compared. They
are less helpful when actual expected cover times need to be computed
for empirical networks where $N$ is fixed and comparative or relative
statements are insufficient.

Unfortunately, a general procedure for estimating the actual cover
time for arbitrary complex networks, as well as the connection between
the mean cover time and FPT observables is lacking. In the following
we present a method that estimates the cover time using passage time
statistics. 

\begin{figure*}
\begin{centering}
\includegraphics[width=0.75\textwidth]{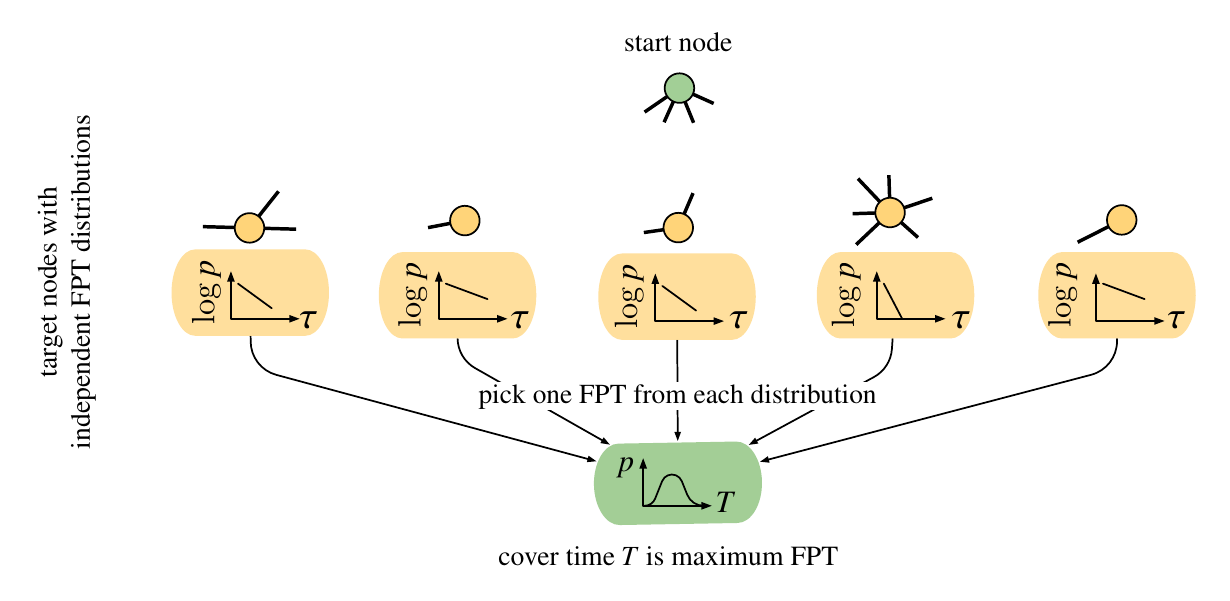}
\par\end{centering}
\caption{Illustration of our approach. Each node of the network is viewed as
an independent entity that can be visited by the walker starting at
the green node. Each target node $v$ has its own first passage time
(FPT) distribution which is asymptotically (for larger times $\tau$)
distributed as $\propto\exp(-\tau/\tau_{v})$ with its global mean
FPT $\tau_{v}$. In order to compute the cover time, we draw one FPT
$\tau$ from every target's distribution. Then the cover time is given
as the maximum time of all drawn FPTs.\label{fig:heuristic}}
\end{figure*}

\subsection{Cover Time}

Recently it has been found that if a random walk process equilibrates
quickly, i.e.~the initial concentration of random walkers approaches
the equilibrium concentration in a small number of time steps $t_{\mathrm{rlx}}\ll N$,
the information about the start node is lost \cite{lau_asymptotic_2010}
and the first passage time at destination $v$ is (for larger times
$\tau$) distributed asymptotically according to
\[
p_{v}(\tau)\propto\exp(-\tau/\tau_{v})
\]
where $\tau_{v}$ is the GMFPT of Eq.~(\ref{eq:GMFPT}). $\tau_{v}$
can differ between nodes and depends on the topological features of
the network only. Note that in the following paragraphs we will often
refer to the FPT decay rate $\beta_{v}=\tau_{v}^{-1}$ instead of
the GMFPT, simplifying the notation.

In order to find the mean cover time from the collection of distributions
$p_{v}(\tau)$ we proceed as illustrated in Fig.~\ref{fig:heuristic}.
Excluding the start node $u$, we pick an FPT $t_{v}$ for each target
node $v$ from their respective FPT distribution $p_{v}$ at random,
resulting in a set of $N-1$ FPT that we call $\mathcal{F}$. Consequently,
the cover time $T_{u}$ is given as the maximum element of $\mathcal{F}$.
In order to find the distribution of this maximum, we compute the
probability that a time $T$ is an upper bound of this set as the
probability that no element of $\mathcal{F}$ is larger than $T$,
yielding
\begin{eqnarray}
P_{u}(T) & = & P_{u}(\mathrm{``}T\ \mathrm{is}\ \mathrm{an\ upper\ bound\ of\ FPTs"})\nonumber \\
 & = & P_{u}(t_{v}\leq T\ \forall t_{v}\in\mathcal{F})=\prod_{v\neq u}P_{u}(t_{v}\leq T).\label{eq:cdf_cover_time_single_node}
\end{eqnarray}
We further approximate our result by assuming a continuous time distribution,
easing the computations without significantly changing the outcome
as explained in App.~\ref{sec:continuous_time}. Then, the probability
that any time $t_{v}\in\mathcal{F}$ is lower than or equal to $T$
is 
\begin{align}
P(t_{v}\leq T) & =\int\limits _{0}^{T}\mathrm{d}t\ p_{v}(t)=\int\limits _{0}^{T}\mathrm{d}t\ \beta_{v}\exp(-\beta_{v}t)\nonumber \\
 & =1-\exp(-\beta_{v}T).\label{eq:cdf_cont_time_result}
\end{align}
 Eqs.~(\ref{eq:cdf_cover_time_single_node}) and (\ref{eq:cdf_cont_time_result})
yield the cumulative distribution function for the cover time, 
\[
P_{u}(T)=\prod_{v\neq u}\Big(1-\exp(-\beta_{v}T)\Big)
\]
and the expected cover time 
\begin{align}
T_{u} & \approx\int\limits _{0}^{\infty}\mathrm{d}T\ \left[1-\prod_{v\neq u}\left(1-\exp(-\beta_{v}T)\right)\right]-\frac{1}{2}.\label{eq:cover_time_integral_single_node}
\end{align}
As discussed in App.~\ref{sec:continuous_time}, the additional shift
of $1/2$ emerges when changing from discrete time to continuous time.
However, since the mean cover time is usually $T_{u}\gg1/2$, we will
omit this shift, introducing relative error of $(2T_{u})^{-1}.$ We
can find the global mean cover time by averaging over target nodes
$u$ as
\begin{align}
\expv T & \approx\int\limits _{0}^{\infty}\mathrm{d}T\left[1-P(T)\frac{1}{N}\sum_{u=1}^{N}\frac{1}{1-\exp(-\beta_{u}T)}\right]\label{eq:integral_mean_cover_time_complicated}
\end{align}
with 
\begin{equation}
P(T)=\prod_{v=1}^{N}\left(1-\exp(-\beta_{v}T)\right).\label{eq:cdf_cover_time}
\end{equation}
\begin{figure*}
\begin{centering}
\includegraphics[width=1\textwidth]{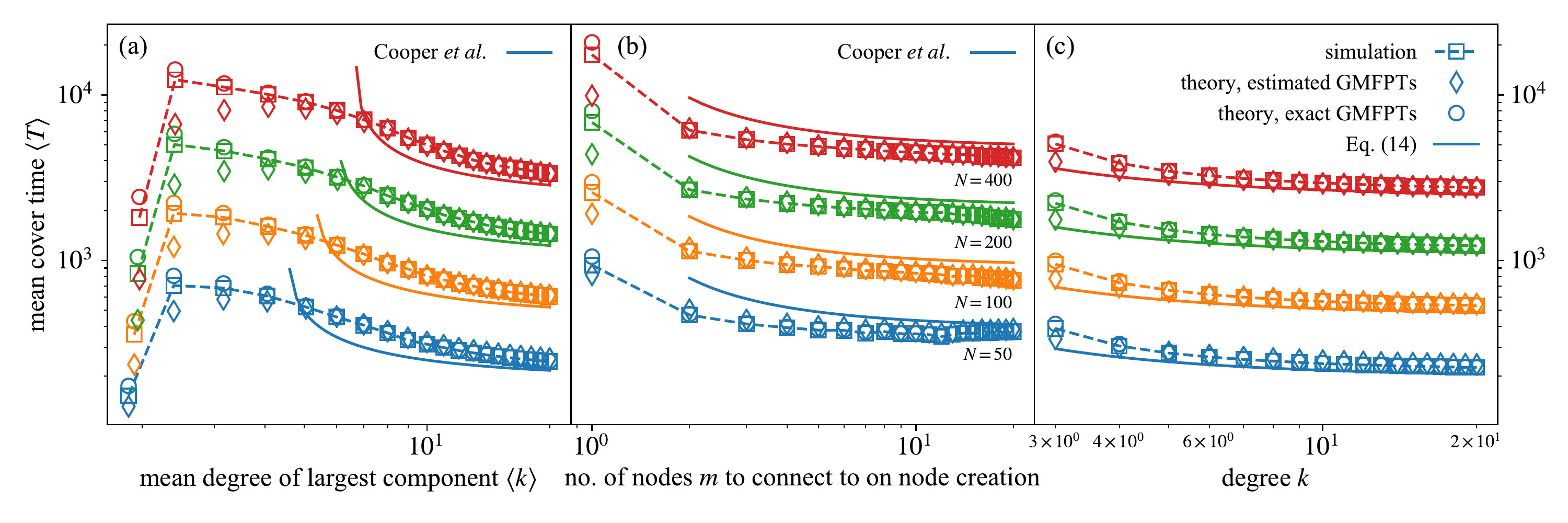}
\par\end{centering}
\centering{}\caption{\label{fig:ER_BA_cover_time}The mean cover time $\expv T$ of the
largest component of (a) Erd\H{o}s\textendash Rényi (ER) networks,
(b) Barabási\textendash Albert (BA) networks , and (c) random $k$-regular
networks. Shown are averages measured from 1000 simulations per data
point ($\boxempty$) and both estimations of the mean cover time using
($\diamondsuit$) estimated GMFPTs from the target nodes' degrees
Eq.~(\ref{eq:T_heur_est}) and ($\varbigcirc$) exact GMFPTs Eq.~(\ref{eq:T_heur_ex})
computed from the unnormalized graph Laplacian's spectrum in (a) and
(b). Respectively, we used Eq.~(\ref{eq:cover_time_equal_degree})
for random $k$-regular networks in (c). Dashed lines are simulation
results and a guide to the eye. Further displayed are the asymptotic
results derived by Cooper, \textit{et. al.} for the ER \cite{cooper_cover_2007}
and BA \cite{cooper_cover_2007-1} network model as well as the lower
bound Eq.~(\ref{eq:cover_time_lower_bound_k_reg_networks}) for random
$k$-regular networks.}
\end{figure*}
However, as shown in App.~\ref{sec:approx-cover-time-integral},
introducing small relative error of order $\mathcal{O}\left((N\log N)^{-1}\right)$
for the networks discussed in this paper, we will make use of a simpler
integral to find 
\begin{align}
\expv{T} & \approx\int\limits _{0}^{\infty}\mathrm{d}T\left[1-P(T)\right]=\int\limits _{0}^{\infty}\mathrm{d}T\left[1-\prod_{v=1}^{N}\left(1-\exp(-\beta_{v}T)\right)\right]\label{eq:cover_time_integral}\\
 & =\sum_{\mathcal{S}\in\mathcal{P}^{*}(\mathcal{V})}(-1)^{|\mathcal{S}|+1}\left(\sum_{v\in\mathcal{S}}\beta_{v}\right)^{-1}.\nonumber 
\end{align}
Here, $\mathcal{V}$ is the set of all nodes and $\mathcal{P}^{*}(\mathcal{V})$
is the set of all possible subsets of $\mathcal{V}$ excluding the
empty set. Conceptually, this integral equals a situation where an
additional node is inserted on which every random walk starts but
which can never be visited again. Even though one can solve integral
Eq.~(\ref{eq:cover_time_integral}) analytically to obtain the result
above, in practice it is more feasible to solve the integral numerically
than iterating over $\mathcal{P}^{*}(\mathcal{V})$ which has $2^{N}-1$
elements and hence becomes very large rather quickly.

\begin{figure*}
\begin{centering}
\includegraphics[width=1\textwidth]{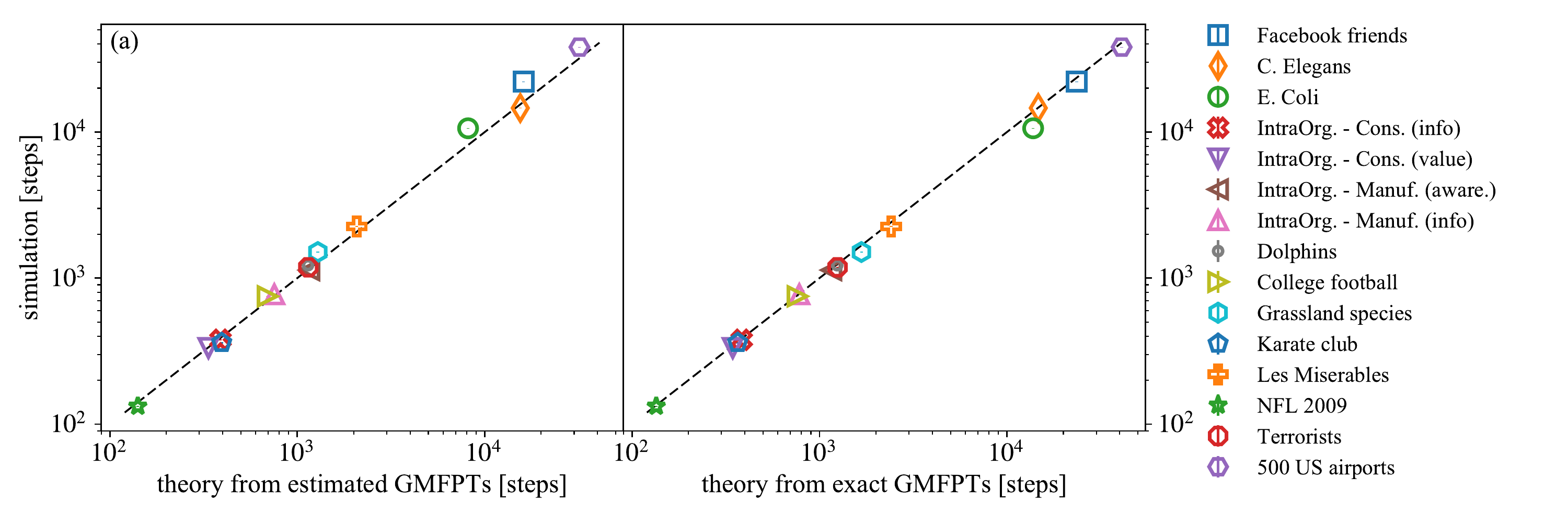}
\par\end{centering}
\begin{centering}
\includegraphics[width=1\textwidth]{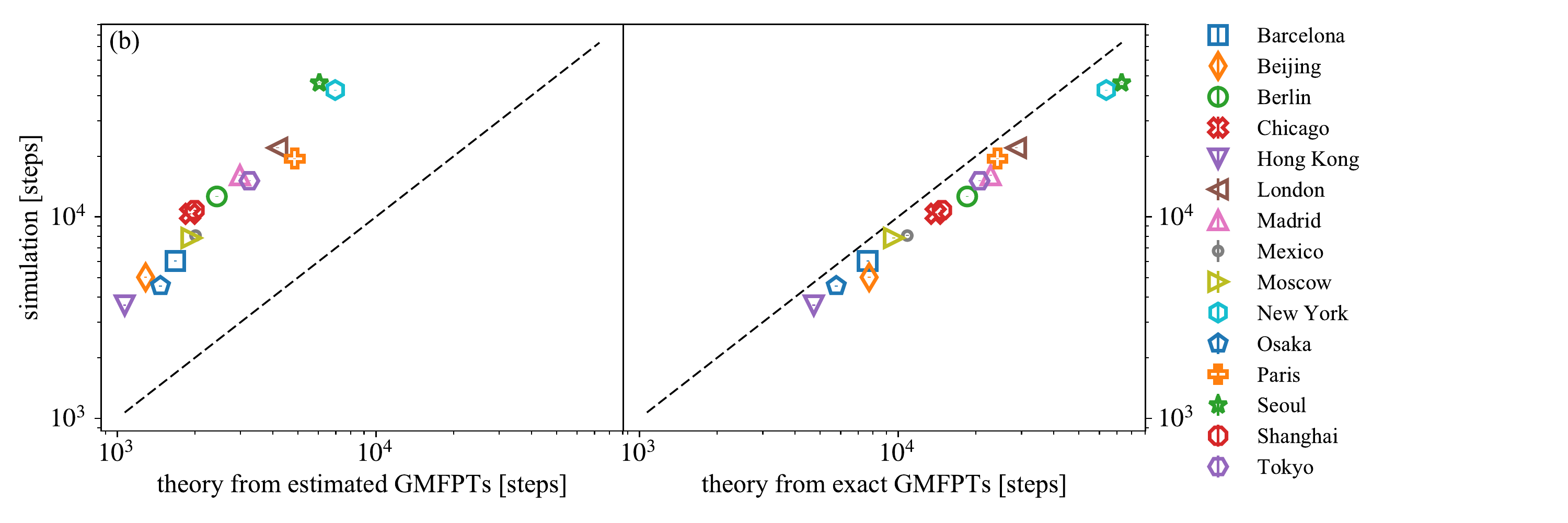}
\par\end{centering}
\caption{Mean cover times of simple discrete time random walks on the largest
component of (a) various real-world networks (data sources and relative
errors given in Tab.~\ref{tab:Mean-cover-times-various-networks})
and (b) various subway networks (data source and relative errors given
in Tab.~\ref{tab:Mean-cover-times-subway-networks}). The estimated
cover times are compared to the measured cover times (from 50 simulations
for each data point). The dashed lines represent the ideal case $\expv{T^{\mathrm{sim}}}=\expv{T^{\mathrm{es/ex}}}$.
Theoretical results are computed from (left) estimated GMFPTs from
the target nodes' degrees Eq.~(\ref{eq:T_heur_est}) and (right)
exact GMFPTs computed from the unnormalized graph Laplacian's spectrum
Eq.~(\ref{eq:T_heur_ex}). \label{fig:Mean-cover-times}}
\end{figure*}
Now, the estimation of the global mean cover time reduces to an efficient
estimation of the FPT decay rates $\beta_{v}$. There are two ways
to estimate the decay rates with the GMFPTs as described in Sec.~\ref{subsec:simple_random_walk}.
Using the estimation of the lower bound Eq.~(\ref{eq:GMFPT_lower_bound}),
the estimated global mean cover time is given by
\begin{equation}
\expv{T^{\mathrm{es}}}\geq\int\limits _{0}^{\infty}\mathrm{d}T\left[1-\prod_{v=1}^{N}\left(1-\exp\left(-Tk_{v}\frac{1-\expv k^{-1}}{N\expv k}\right)\right)\right].\label{eq:T_heur_est}
\end{equation}
The advantage of this method is that only the network's degree sequence
$k_{v}$ needs to be known in order to estimate the global mean cover
time. However, this method can obviously only account for a lower
bound. We can also compute the exact GMFPTs using Eq.~(\ref{eq:GMFPT_Laplacian}).
In this case the computed global mean cover time is 
\begin{equation}
\expv{T^{\mathrm{ex}}}=\int\limits _{0}^{\infty}\mathrm{d}T\left[1-\prod_{v=1}^{N}\left(1-\exp\left(-\frac{T}{\tau_{v}^{\mathrm{ex}}}\right)\right)\right].\label{eq:T_heur_ex}
\end{equation}
\begin{table*}[t]
\centering{}\begin{tabular}{lrrrrrrr}
\hline\hline
 Network                                                                            &   $N$ &   $\left\langle k\right\rangle$ &   $\frac{\left\langle T^\mathrm{sim}\right\rangle}{N\log N}$ &   $\frac{\left\langle T^\mathrm{es}\right\rangle}{N\log N}$ &   $\left|1-\frac{\left\langle T^\mathrm{sim}\right\rangle}{\left\langle T^\mathrm{es}\right\rangle}\right|$ &   $\frac{\left\langle T^\mathrm{ex}\right\rangle}{N\log N}$ &   $\left|1-\frac{\left\langle T^\mathrm{sim}\right\rangle}{\left\langle T^\mathrm{ex}\right\rangle}\right|$ \\
\hline
 First author's Facebook friends network \cite{maier_b.f._2017}                     &   329 &                            11.9 &                                                        11.61 &                                                        8.45 &                                                                                                        0.37 &                                                       12.36 &                                                                                                       0.061 \\
 C. Elegans neural network \cite{watts_collective_1998}                             &   297 &                            14.6 &                                                         8.64 &                                                        9.15 &                                                                                                        0.06 &                                                        8.69 &                                                                                                       0.006 \\
 E. Coli protein interaction \cite{shen-orr_network_2002}                           &   329 &                             2.8 &                                                         5.57 &                                                        4.27 &                                                                                                        0.30 &                                                        7.24 &                                                                                                       0.231 \\
 Intra-org. contacts - Cons. (info) \cite{cross_hidden_2004}                        &    43 &                            15.3 &                                                         2.34 &                                                        2.41 &                                                                                                        0.03 &                                                        2.38 &                                                                                                       0.018 \\
 Intra-org. contacts - Cons. (value)                                                &    44 &                            16.0 &                                                         2.00 &                                                        2.02 &                                                                                                        0.01 &                                                        2.07 &                                                                                                       0.036 \\
 Intra-org. contacts - Manuf. (awareness)                                           &    77 &                            25.5 &                                                         3.39 &                                                        3.47 &                                                                                                        0.02 &                                                        3.46 &                                                                                                       0.021 \\
 Intra-org. contacts - Manuf. (info)                                                &    76 &                            23.3 &                                                         2.35 &                                                        2.29 &                                                                                                        0.03 &                                                        2.37 &                                                                                                       0.009 \\
 Social interaction in dolphins \cite{lusseau_bottlenose_2003}                      &    62 &                             5.1 &                                                         4.79 &                                                        4.46 &                                                                                                        0.07 &                                                        4.86 &                                                                                                       0.015 \\
 American college football \cite{girvan_community_2002}                             &   115 &                            10.7 &                                                         1.37 &                                                        1.27 &                                                                                                        0.08 &                                                        1.40 &                                                                                                       0.017 \\
 Food web of grassland species \cite{dawah_structure_1995}                          &    75 &                             3.0 &                                                         4.66 &                                                        3.97 &                                                                                                        0.17 &                                                        5.17 &                                                                                                       0.099 \\
 Zachary's Karate club \cite{zachary_information_1977}                              &    34 &                             4.5 &                                                         3.01 &                                                        3.29 &                                                                                                        0.09 &                                                        3.05 &                                                                                                       0.015 \\
 Interactions in ``Les Mis\'erables'' \cite{knuth_stanford_1993}                    &    77 &                             6.6 &                                                         6.75 &                                                        6.21 &                                                                                                        0.09 &                                                        7.21 &                                                                                                       0.063 \\
 Matches of the NFL 2009 \cite{aicher_learning_2015}                                &    32 &                            13.2 &                                                         1.20 &                                                        1.27 &                                                                                                        0.06 &                                                        1.21 &                                                                                                       0.015 \\
 Network of associations between terrorists \cite{krebs_mapping_2002}               &    62 &                             4.9 &                                                         4.63 &                                                        4.47 &                                                                                                        0.04 &                                                        4.87 &                                                                                                       0.049 \\
 Connections between 500 largest US airports \cite{colizza_reaction-diffusion_2007} &   500 &                            11.9 &                                                        12.29 &                                                       10.30 &                                                                                                        0.19 &                                                       13.18 &                                                                                                       0.067 \\
\hline
\end{tabular}\caption{Mean cover times of simple discrete time random walks on the largest
component of various networks, in units of the cover time on a complete
graph with equal node count. Displayed is the number of nodes $N$,
the mean degree $\expv k$ of the largest component and the measured
mean cover time $\expv{T^{\mathrm{sim}}}$ extracted from 50 simulations
per network with one walker starting on every node. Additionally shown
are both theoretical estimations of the cover time using (a) estimated
GMFPTs $\expv{T^{\mathrm{es}}}$ from the target nodes' degrees, Eq.~(\ref{eq:T_heur_est})
and (b) exact GMFPTs $\expv{T^{\mathrm{ex}}}$ computed from the unnormalized
graph Laplacian's spectrum Eq.~(\ref{eq:T_heur_ex}). Both estimations
are given with their relative error to the simulated mean cover time.
Note that all networks have been symmetrized and an unweighted link
$(u,v)$ has been created if a weight between two nodes was $w_{uv}>0$.
For the Intra-organizational networks, we created a link if both nodes
put something else than ``I do not know this person'' in their questionnaire.
These values are additionally shown in Fig.~\ref{fig:Mean-cover-times}
(top).\label{tab:Mean-cover-times-various-networks}}
\end{table*}

\subsection{Cover time of networks with equal GMFPTs}

Let us consider a network in which all nodes have approximately the
same GMFPT $\tilde{\tau}$ and on which a random walk equilibrates
quickly ($t_{\mathrm{rlx}}\ll N$) such that we can estimate the mean
cover time using Eq.~(\ref{eq:cover_time_integral_single_node}).
We find
\begin{align}
\expv T(\tilde{\tau}) & \approx\int\limits _{0}^{\infty}\mathrm{d}T\ \left[1-\left(1-\exp(-T/\tilde{\tau})\right)^{N-1}\right]\nonumber \\
 & =\tilde{\tau}\Big[\gamma+\psi(N)\Big],\label{eq:cover_time_equal_degree}
\end{align}
 where $\gamma\approx0.57722$ is the Euler-Mascheroni constant, $\psi(z)=\Gamma'(z)/\Gamma(z)$
and $\Gamma(z)$ the gamma function.

An example for networks fulfilling the conditions above are random
$k$-regular networks where all nodes have identical degree and the
networks possess random structure (as opposed to, e.g.~lattice networks
on a torus, where all nodes have identical degree but are only connected
to their nearest neighbors). This includes, e.g.~the complete graph.
The cover time of the complete graph is given as $\expv T=(N-1)\left(\log(N-1)+\gamma+\mathcal{O}(N^{-1})\right)$
\cite{lovasz_random_1996}, a result which is reproduced by Eq.~(\ref{eq:cover_time_equal_degree})
since the GMFPT for each node is $\tilde{\tau}=N-1$ (see App.~\ref{app:gmfpt_comp_graph})
and $\psi(N+1)=\log N+\mathcal{O}(N^{-1})$. For general random $k$-regular
networks, we can use Eq.~(\ref{eq:cover_time_equal_degree}) to find
an approximate scaling relation for the lower bound 
\begin{equation}
\expv T\apprge\frac{k}{k-1}N\log N\label{eq:cover_time_lower_bound_k_reg_networks}
\end{equation}
using the GMFPT lower bound Eq.~(\ref{eq:GMFPT_lower_bound}), the
fact that $k_{v}=\expv k=k$ and $\psi(N+1)=\log N+\mathcal{O}(N^{-1})$.

\section{Results}

We compared the predictions of Eqs.~(\ref{eq:T_heur_est}) and (\ref{eq:T_heur_ex})
with simulation results for single component ER, BA and real-world
networks, as well as Eq.~(\ref{eq:cover_time_equal_degree}) for
random $k$-regular networks. On every node we placed a walker at
time $t=0$. Subsequently, we let each walker do a random walk as
described in Sec.~\ref{subsec:simple_random_walk}. Each walker proceeded
until it visited each node at least once, completing total coverage
and marking cover time $T_{u}$. $\expv T$ was computed as the average
of all $T_{u}$. For a more detailed description of the numerical
methods as well as the used code, see App.~\ref{app:sim_and_num}.

For both ER and BA networks, we generated networks with $N\in\{50,100,200,400\}$
nodes, ER networks with node connection probability $\{k/(N-1):k\in\mathbb{N},\ 1\leq k\leq20\}$,
and BA networks with $\{m:m\in\mathbb{N},\ 1\leq m\leq20\}$, $m$
representing the number of new links per node at creation. In order
to test Eq.~(\ref{eq:cover_time_equal_degree}), we generated random
$k$-regular networks using the algorithm given in \cite{steger_generating_1999}
with $N\in\{50,100,200,400\}$ nodes and node degree $k_{u}=k\ \forall\,u\in\mathcal{V}$,
scanning integer degrees $\{k:k\in\mathbb{N},\ 3\leq k\leq20\}$.
After creating each network, we extracted the largest component, ran
discrete time random walks as described above and estimated the cover
time using Eqs.~(\ref{eq:T_heur_est}), (\ref{eq:T_heur_ex}) and
Eq.~(\ref{eq:cover_time_equal_degree}), respectively, for 1000 networks
each. For Eq.~(\ref{eq:cover_time_equal_degree}) and the random
$k$-regular networks, we computed $\tilde{\tau}^{\mathrm{es}}=N/(1-k^{-1})$
and $\tilde{\tau}^{\mathrm{ex}}=N^{-1}\sum_{v=1}^{N}\tau_{v}^{\mathrm{ex}}$,
respectively.

The theoretic results are in agreement with the simulation results,
as can be seen in Fig.~\ref{fig:ER_BA_cover_time}. The relative
error decreases with increasing number of nodes $N$ as well as increasing
mean degree $\expv k$ and quickly reaches values below 1\%. Unsurprisingly,
our method performs better compared to the results of \cite{cooper_cover_2007,cooper_cover_2007-1}
due to the asymptotic nature of the latter.

We furthermore simulated random walks on the largest component of
15 real-world networks, listed in Tab.~\ref{tab:Mean-cover-times-various-networks}.
Inititially directed networks were converted to undirected networks
replacing every directed link with an undirected link. For weighted
networks we assigned an undirected link $(u,v)$ if a weight was $w_{uv}>0$.
For the intra-organizational networks \cite{cross_hidden_2004}, employees
had to fill out questionnaires regarding their relationships to co-workers.
Here, we assigned an undirected link $(u,v)$ if both $u$ and $v$
marked something else than ``I do not know this person''. As can
be seen in Tab.~\ref{tab:Mean-cover-times-various-networks}, our
method produces results that are very close to the simulated values
(mostly relative errors of $<10\%$). Exceptions are the computed
cover times for the E. coli protein interaction network \cite{shen-orr_network_2002}
with a relatively high relative error of $\approx23\%$ and the grassland
food web \cite{dawah_structure_1995} with a relative error of $\approx10\%$.

\begin{figure}
\includegraphics[width=1\columnwidth]{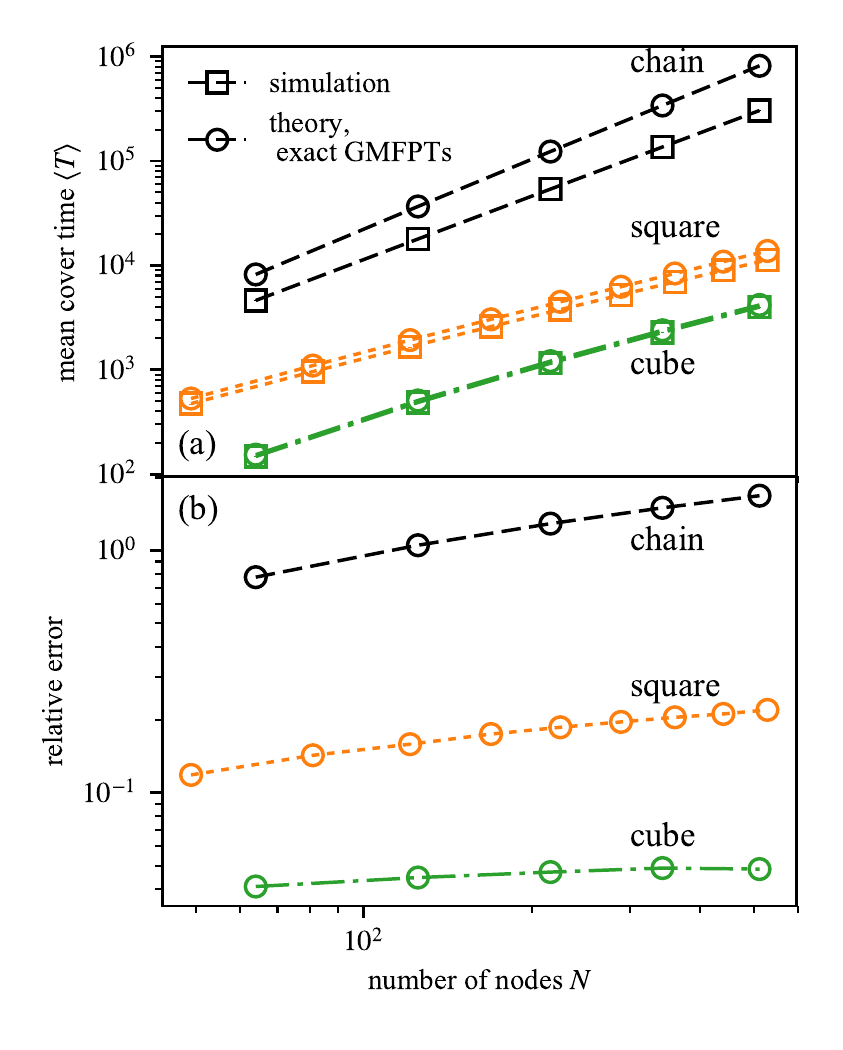}\caption{\label{fig:lattice}Example of our method yielding results with rather
large deviations from simulations.(a) Mean cover time for low-dimensional
$(d\leq2)$ lattices as well as lattices in dimension $d=3$ as ($\boxempty$)
an average over 1000 simulations for each data point and ($\varbigcirc$)
theoretical result from exact GMFPTs computed from the unnormalized
graph Laplacian's spectrum Eq.~(\ref{eq:T_heur_ex}). (b) The relative
error is is increasing with increasing system size but is comparably
lower for $d=3$ (cubes).}
\end{figure}
\begin{table*}[t]
\centering{}\begin{tabular}{lrrrrrrr}
\hline\hline
 Network   &   $N$ &   $\left\langle k\right\rangle$ &   $\frac{\left\langle T^\mathrm{sim}\right\rangle}{N\log N}$ &   $\frac{\left\langle T^\mathrm{es}\right\rangle}{N\log N}$ &   $\left|1-\frac{\left\langle T^\mathrm{sim}\right\rangle}{\left\langle T^\mathrm{es}\right\rangle}\right|$ &   $\frac{\left\langle T^\mathrm{ex}\right\rangle}{N\log N}$ &   $\left|1-\frac{\left\langle T^\mathrm{sim}\right\rangle}{\left\langle T^\mathrm{ex}\right\rangle}\right|$ \\
\hline
 Barcelona &   128 &                             2.2 &                                                          9.7 &                                                        2.70 &                                                                                                         2.6 &                                                       12.29 &                                                                                                        0.21 \\
 Beijing   &   104 &                             2.2 &                                                         10.4 &                                                        2.67 &                                                                                                         2.9 &                                                       16.00 &                                                                                                        0.35 \\
 Berlin    &   170 &                             2.1 &                                                         14.4 &                                                        2.78 &                                                                                                         4.2 &                                                       21.19 &                                                                                                        0.32 \\
 Chicago   &   141 &                             2.1 &                                                         14.8 &                                                        2.74 &                                                                                                         4.4 &                                                       20.02 &                                                                                                        0.26 \\
 Hong Kong &    82 &                             2.1 &                                                         10.1 &                                                        2.96 &                                                                                                         2.4 &                                                       13.07 &                                                                                                        0.23 \\
 London    &   266 &                             2.3 &                                                         14.8 &                                                        2.80 &                                                                                                         4.3 &                                                       19.26 &                                                                                                        0.23 \\
 Madrid    &   209 &                             2.3 &                                                         14.4 &                                                        2.67 &                                                                                                         4.4 &                                                       20.42 &                                                                                                        0.29 \\
 Mexico    &   147 &                             2.2 &                                                         11.0 &                                                        2.74 &                                                                                                         3.0 &                                                       14.80 &                                                                                                        0.26 \\
 Moscow    &   134 &                             2.3 &                                                         12.0 &                                                        2.95 &                                                                                                         3.1 &                                                       14.66 &                                                                                                        0.18 \\
 New York  &   433 &                             2.2 &                                                         16.2 &                                                        2.65 &                                                                                                         5.1 &                                                       24.25 &                                                                                                        0.33 \\
 Osaka     &   108 &                             2.3 &                                                          9.0 &                                                        2.91 &                                                                                                         2.1 &                                                       11.40 &                                                                                                        0.21 \\
 Paris     &   299 &                             2.4 &                                                         11.4 &                                                        2.85 &                                                                                                         3.0 &                                                       14.22 &                                                                                                        0.20 \\
 Seoul     &   392 &                             2.2 &                                                         19.7 &                                                        2.58 &                                                                                                         6.6 &                                                       31.24 &                                                                                                        0.37 \\
 Shanghai  &   148 &                             2.1 &                                                         14.6 &                                                        2.69 &                                                                                                         4.4 &                                                       19.96 &                                                                                                        0.27 \\
 Tokyo     &   217 &                             2.4 &                                                         12.9 &                                                        2.78 &                                                                                                         3.6 &                                                       17.69 &                                                                                                        0.27 \\
\hline
\end{tabular}\caption{Same procedure as in Tab.~\ref{tab:Mean-cover-times-various-networks},
but for subway networks of big cities, taken for the year 2009 from
\cite{roth_long-time_2012}. These values are additionally shown in
Fig.~\ref{fig:Mean-cover-times} (bottom).\label{tab:Mean-cover-times-subway-networks}}
\end{table*}
Additionally, we performed simulations on $d$-dimensional lattices
of dimension $d\in\{1,2,3\}$ (chains, squares and cubes) using node
numbers $N\in\{(2n+1)^{2}:n\in\mathbb{N},\ 2\leq n\leq12\}$ for $d=2$
and $N\in\{n^{3}:n\in\mathbb{N},4\leq n\leq8\}$ for $d=1$ and $d=3$.
For low-dimensional lattice networks with $d\leq2$, the relaxation
time is large compared to a variety of complex networks (see Fig.~1
in \cite{elsasser_tight_2011}). Hence, we suspect that our method
will not perform well for low-dimensional lattice networks and networks
where nodes are embedded in low-dimensional space at position $\boldsymbol{r}_{u}$
with short-range connection probability $\pi_{vu}\propto r_{vu}^{-\omega}$
with $r_{vu}=\left|\boldsymbol{r}_{u}-\boldsymbol{r}_{v}\right|$
and $\omega>d$ as those networks are comparable to low-dimensional
lattices concerning search processes \cite{kleinberg_small-world_2000}.
Indeed, as can be seen in Fig.~\ref{fig:lattice}, the relative error
between simulation and heuristic results increases with increasing
$N$, up to $\approx110\%$ for chains and $\approx10\%$ for square
lattices using exact GMFPTs, whereas smaller relative errors of up
to $\approx4\%$ are reached for cube lattices. Similar results are
obtained for real-world networks embedded in a two-dimensional space
with short-range connection probability such as subway networks \cite{roth_long-time_2012}
(shown in Tab.~\ref{tab:Mean-cover-times-subway-networks} and Fig.~\ref{fig:Mean-cover-times}).
Here, the estimation from estimated GMFPTs systematically underestimates
the cover time while using exact GMFPTs yields an overestimation of
the cover time by $\approx20\%-40\%$.

Generally, the more exact result of GMFPTs calculated via the unnormalized
graph Laplacian gives results with lower relative error than using
lower bound GMFPTs, as expected.

\begin{figure}
\includegraphics[width=1\columnwidth]{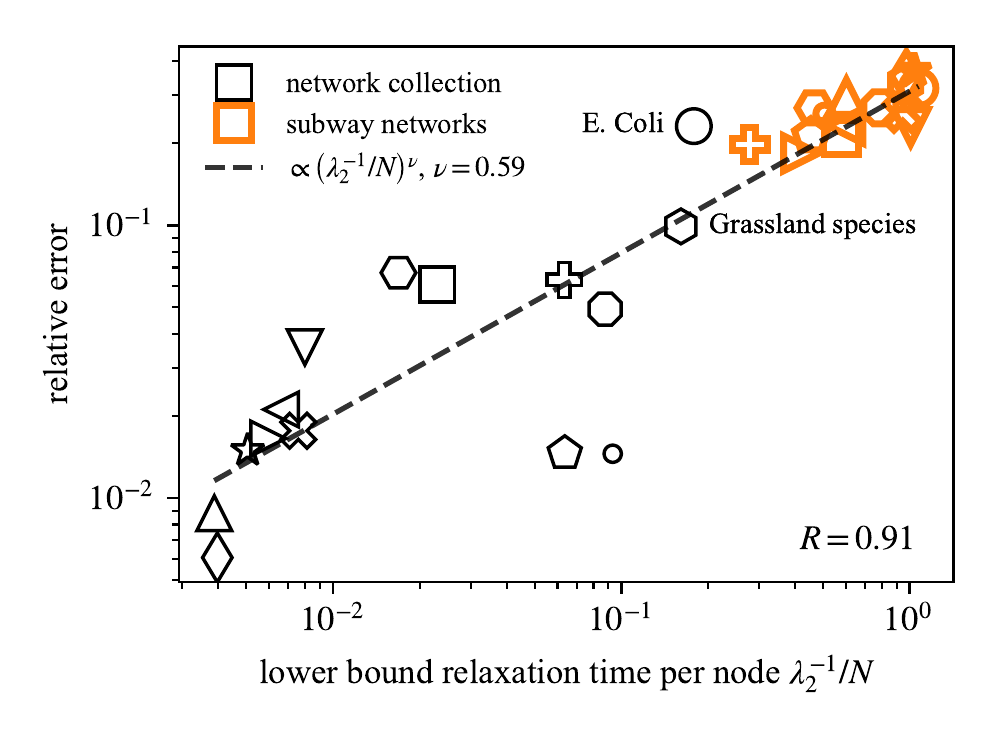}\caption{\label{fig:rel_err_vs_relax_time}The relative error of our heuristic
is increasing with increasing inverse second smallest eigenvalue $\lambda_{2}^{-1}$
of the unnormalized graph Laplacian, indicating that our method produces
higher deviations for networks with relatively high relaxation times
as per Eq.~(\ref{eq:t_rlx}). Networks are marked with the same symbols
as in Fig.~\ref{fig:Mean-cover-times}.}
\end{figure}

Concerning the impact of network structure on the error of our heuristic
compared to the true mean cover time, we found that a large relaxation
time directly influences the relative error. Since we derived our
results under the assumption that the relaxation time is $t_{\mathrm{rlx}}\ll N$,
we measured relative error against the ratio $t_{\mathrm{rlx}}/N$
and used Eq.~(\ref{eq:t_rlx}) to find
\[
\mathrm{rel.\ err.}\propto\left(\frac{\lambda_{2}^{-1}}{N}\right)^{-0.59}
\]
as can be seen in Fig.~\ref{fig:rel_err_vs_relax_time} indicating
that increasing relaxation time increases the error of our heuristic.

\section{Conclusions}

We studied the cover time of simple discrete time random walks on
single component complex networks with $N$ nodes. Treating each target
node as independent from the start node, we were able to find the
cumulative distribution function of the cover time by finding the
maximum of $N$ drawn FPTs from the $N$ target nodes' FPT distributions
which solely depend on their GMFPT. Using this method, the complexity
of finding the mean cover time of an arbitrary complex network is
heavily decreased since the problem is practically reduced to finding
the nodes' GMFPTs using simple estimations or spectral methods, which
is computationally much more feasible than simulating random walks
starting on every node, especially for large networks.

We showed that this procedure yields reliable estimations of the mean
cover time for a variety of networks where random walks decay quickly,
namely ER networks, BA networks, random $k$-regular networks, and
a collection of real-world networks. We furthermore showed that for
low-dimensional lattices as well as networks which are embedded in
low-dimensional space and have short-range connection probability,
e.g.~subway networks, our method does not produce reliable results.
We were able to map this deviation of the heuristic result to the
ratio of relaxation time per node, indicating that for networks with
high relaxation time the heuristic will produce more erroneous results.
The large deviations for the E. Coli interaction network and the grassland
species food web is most likely related to the fact that those networks
are strongly hierarchically clustered \cite{clauset_hierarchical_2008,ravasz_hierarchical_2002}
and hence similar to low-dimensional spatial networks with short-range
interaction probability concerning random walks and search processes
\cite{watts_identity_2002}. The exact influence of a strong hierarchically
organized network structure on cover time is, however, a task for
future investigations.

Finally, note that even though we derived our results for unweighted
networks, they can be easily made applicable to weighted networks
and subclasses of directed networks, as they only depend on the calculation
of GMFPTs. Those, in turn, depend solely on the transition matrix
$W_{vu}=A_{vu}/\sum_{u}A_{vu}$ which is similarly defined for weighted
and directed networks.
\begin{acknowledgments}
We would like to thank the reviewers for their detailed and helpful
comments that lead to a considerable revision of our manuscript and
additions of detailed error estimations. We also want to thank M.
Saftzystidling and F. Ackerling for inspirational discussions.
\end{acknowledgments}

\appendix

\section{The GMFPT of a complete graph}

\label{app:gmfpt_comp_graph}Suppose a random walker starts at any
node $u$. The probability to reach any other node of the network
in one time step is $p=1/(N-1)$. Looking at a single target node
$v$ we want to calculate the probability that $v$ is first passaged
at time $t$, which is given as 
\[
p_{t}=(1-p)^{t-1}p.
\]
Hence, the GMFPT for every target node is 
\[
\tilde{\tau}=\sum_{t=1}^{\infty}tp_{t}=-p\frac{\partial}{\partial p}\frac{1}{p}=\frac{1}{p}=N-1.
\]

\section{Continuous time approximation}

\label{sec:continuous_time}Since we are investigating discrete time
random walks in this study, the probability distributions are actually
probability mass functions (pmfs) and expectation values should be
calculated using series instead of integrals. In the following, we
discuss differences and introduced errors by using continuous distributions
instead.

First, the discrete time cumulative distribution function for first
passage time $\tau$ at node $v$ is calculated using pmf 
\begin{align*}
p_{v,\tau} & =\exp(-\beta_{v}\tau)\Bigg/\sum_{t=1}^{\infty}\exp(-\beta_{v}t)\\
 & =[1-\exp(-\beta_{v})]\exp(-\beta_{v}(\tau-1)),
\end{align*}
yielding 
\begin{align*}
P(t_{v}\leq T) & =\sum_{t=1}^{T}p_{v,t}=(1-\exp(-\beta_{v}))\sum_{t=1}^{T}\exp(-\beta_{v}(t-1))\\
 & =1-\exp(-\beta_{v}T)
\end{align*}
which is equal to the continuous time result in Eq.~(\ref{eq:cdf_cont_time_result}).
The mean cover time is then given as the series, respectively partial
sum
\[
T_{u}=\sum_{T=1}^{\infty}\left[1-P_{u}(T)\right]=\sum_{T=1}^{\infty}\bar{P}_{u}(T)\approx\sum_{T=1}^{T_{\max}}\bar{P}_{u}(T)
\]
where we approximated the upper boundary using a $T_{\max}$ with
$\bar{P}_{u}(T_{\max})\leq10^{-10}$ (see App.~\ref{app:sim_and_num}).
This partial sum is equal to the trapezoidal approximation of the
integral 
\[
\int\limits _{0}^{T_{\max}}\mathrm{d}T\ \bar{P}_{u}(T)=\frac{1}{2}\bar{P}_{u}(0)+\sum_{T=1}^{T\max-1}\bar{P}_{u}(T)+\frac{1}{2}\bar{P}_{u}(T_{\max})+\Phi
\]
with $\Delta T=1$. Since the function $\bar{P}_{u}(T)$ has value
$\bar{P}_{u}(0)=1$, using the integral instead of the sum introduces
a systematic error of $1/2$. Using the first derivative $\bar{P}'_{u}(T)$,
the error $\Phi$ emerging from the trapezoidal rule can be asymptotically
estimated to be 
\[
|\Phi|=\frac{\Delta T^{2}}{12}\left|\bar{P}'_{u}(T_{\max})-\bar{P}'_{u}(0)\right|
\]
 for $T_{\max}\rightarrow\infty$ \cite{atkinson_introduction_1989}.
With 
\[
\bar{P}'_{u}(T)=-\sum_{v\neq u}\beta_{v}\exp(-\beta_{v}T)\prod_{w\neq v\neq u}\left[1-\exp(-\beta_{w}T)\right]
\]
 we have $\bar{P}'_{u}(0)=0$. In another way, analogous to Eq.~(\ref{eq:cover_time_integral})
we find 
\begin{eqnarray*}
\bar{P}_{u}(T) & = & \sum_{\mathcal{S\in P}^{*}(\mathcal{V}\backslash\{u\})}(-1)^{|\mathcal{S}|+1}\exp\left(-\sum_{v\in\mathcal{S}}\beta_{v}T\right)\\
\left|\bar{P}'_{u}(T)\right| & = & \sum_{\mathcal{S\in P}^{*}(\mathcal{V}\backslash\{u\})}(-1)^{|\mathcal{S}|+1}\left(\sum_{v\in\mathcal{S}}\beta_{v}\right)\exp\left(-\sum_{v\in\mathcal{S}}\beta_{v}T\right).
\end{eqnarray*}
For most nodes the decay rates are $\beta_{v}\lessapprox N^{-1}$
with $\lessapprox$ meaning ``lower or of similar order''. Then
$\sum_{v\in\mathcal{S}}\beta_{v}\lessapprox1$ and hence $\left|\bar{P}'_{u}\right|\lessapprox\bar{P}_{u}$
such that one can safely assume $\left|\bar{P}'_{u}(T_{\max})\right|\lessapprox\bar{P}_{u}(T_{\max})\leq10^{-10}$
yielding absolute error
\[
|\Phi|\lessapprox10^{-11}.
\]

\section{Approximation of mean cover time integral}

\label{sec:approx-cover-time-integral}
\begin{figure}
\includegraphics[width=1\columnwidth]{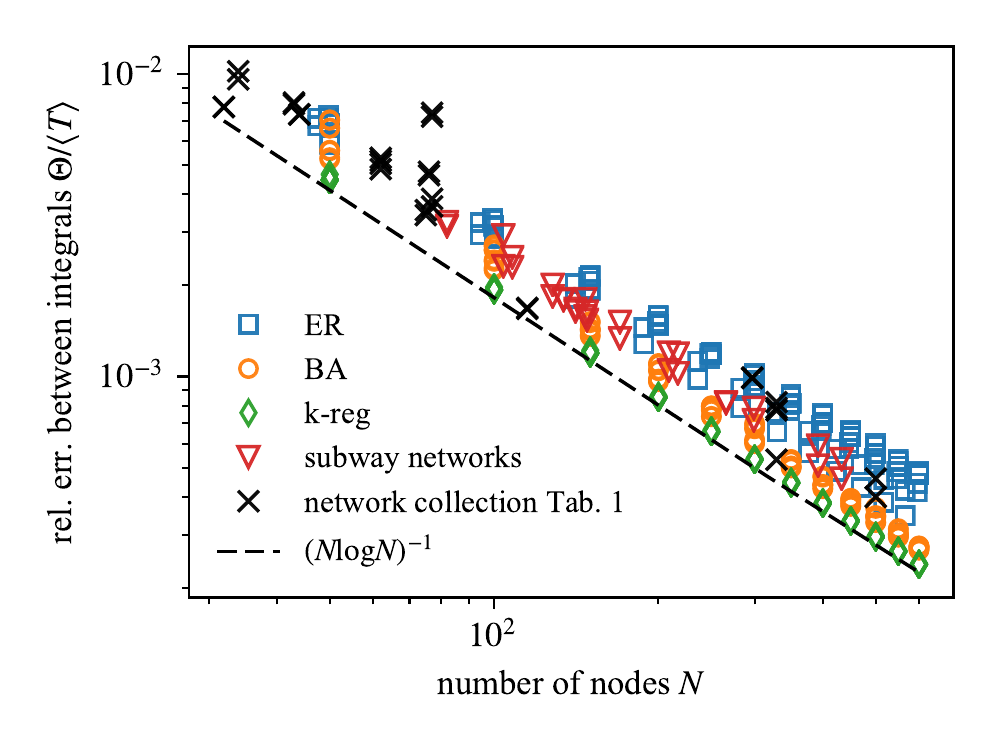}\caption{\label{fig:rel_err_est}Relative error Eq.~(\ref{eq:err_cover_time_integral})
between integrals Eq.~(\ref{eq:integral_mean_cover_time_complicated})
and Eq.~(\ref{eq:cover_time_integral}) for all networks investigated
in this study besides lattices. For each network we show the relative
error for both sets of rates, $\tau_{v}^{\mathrm{es}}$ and $\tau_{v}^{\mathrm{ex}}$.
For ER, BA and random $k$-regular networks, we built means over the
largest components of 100 network realizations for $\expv k\in\{3,5,7,9\}$
and $m\in\{3,5,7,9\}$, respectively. The measured relative errors
are roughly following the scaling relation Eq.~(\ref{eq:err_scaling}).}
\end{figure}

In the following, we show that
instead of solving integral Eq.~(\ref{eq:integral_mean_cover_time_complicated}),
one can safely use Eq.~(\ref{eq:cover_time_integral}). We do so
by calculating the total difference between both as
\begin{align}
\Theta & =\int\limits _{0}^{\infty}\mathrm{d}T\left[1-P(T)\right]-\nonumber \\
 & \qquad-\int\limits _{0}^{\infty}\mathrm{d}T\left[1-P(T)\frac{1}{N}\sum_{v=1}^{N}\frac{1}{1-\exp(-\beta_{v}T)}\right]\nonumber \\
 & =\int\limits _{0}^{\infty}\mathrm{d}T\ P(T)\left[Q(T)-1\right],\label{eq:err_cover_time_integral}
\end{align}
defining $Q(T)=\frac{1}{N}\sum_{v=1}^{N}\left(1-\exp(-\beta_{v}T)\right)^{-1}$.
Note that the cover time cdf $P(T)$ is given by Eq.~(\ref{eq:cdf_cover_time}),
s.t.~both
\begin{align*}
\lim_{T\rightarrow0}P(T) & =0\\
\lim_{T\rightarrow0}P(T)Q(T) & =0
\end{align*}
and
\begin{align*}
\lim_{T\rightarrow\infty}P(T) & =1\\
\lim_{T\rightarrow\infty}P(T)Q(T) & =1,
\end{align*}
meaning that for both integration limits, the integrand approaches
0. In the following we assume that the distribution of decay rates
is relatively homogeneous in the region of small rates, implying that
there is a low number $1<n\ll N$ of nodes $i\in\mathcal{V}_{\mathrm{small}}$
with $n=|\mathcal{V}_{\mathrm{small}}|$ that are of the same order
as $\beta_{\min}=\min\{\beta_{v}:v\in\mathcal{V}\}$. This is a relatively
safe assumption for most network models and real-world networks as
in most cases there are more nodes with small degree (hence small
decay rates) than nodes with high degree (hence high decay rates).
Now suppose the integration approaches a time where $T\approx\beta_{\min}^{-1}$,
implying that, while most terms $1-\exp(-T\beta_{v\notin\mathcal{V}_{\mathrm{small}}})$
are virtually equal to 1 there are still $n$ terms $1-\exp(-T\beta_{i})<1$,
such that $P(T)\approx\prod_{i\in\mathcal{V}_{\mathrm{small}}}^{n}(1-\exp(-T\beta_{i}))\ll1.$
Furthermore, there will already be a majority of terms $1-\exp(-T\beta_{v})\rightarrow1$
which leads to $Q(T)$ approaching $Q(T)\stackrel{N\gg1}{\longrightarrow}1$.
Hence, we can safely assume that for a network with a larger number
of nodes the integrand approaches zero at all times while the global
mean cover time grows quickly and thus the relative error of Eq.~(\ref{eq:cover_time_integral})
is approaching 
\[
\frac{\Theta}{\expv T}\stackrel{N\gg1}{\longrightarrow}0.
\]
In particular, we can calculate the error between the integrals for
random networks with constant GMFPT $\tilde{\tau}$ for every node,
which is given as 
\begin{eqnarray*}
\tilde{\Theta} & = & \tilde{\tau}_{N+1}\Big[\gamma+\psi(N+1)\Big]-\tilde{\tau}_{N}\Big[\gamma+\psi(N)\Big]\\
 & \approx & \tilde{\tau}_{N}\log\left(\frac{N}{N-1}\right)\approx\tilde{\tau}_{N}\frac{1}{N}
\end{eqnarray*}
Where we used $\psi(N+1)=\log N+\mathcal{O}(N^{-1})$ and assumed
$\tilde{\tau}_{N}\sim\tilde{\tau}_{N+1}$. Consequently, we can find
the relative error to be approximately 
\begin{equation}
\frac{\tilde{\Theta}}{\expv T}\approx\frac{1}{N\gamma+N\log(N-1)}\sim\frac{1}{N\log N}.\label{eq:err_scaling}
\end{equation}
Even though this relation is derived for the special case of networks
where every node has the same GMFPT, a numerical analysis of Eq.~(\ref{eq:err_cover_time_integral})
reveals that this scaling relation holds approximately for all networks
investigated in this study, as can be seen in Fig.~(\ref{fig:rel_err_est}).

\section{Simulations and numerical evaluations of the mean cover time}

\label{app:sim_and_num}The code used for the random walk simulations
is a standard implementation of discrete time random walks on networks
and available online as a C++/Python/Matlab package, see \cite{maier_cnetworkdiff_2017}.

In order to evaluate the mean cover time, the integrals Eqs.~(\ref{eq:T_heur_est})
and (\ref{eq:T_heur_ex}) were solved numerically. Note that for the
numerical integration, an upper integration bound of infinity can
be problematic when the decay region of the integrand is unknown.
Hence, we chose an upper integration limit $T_{\max}$ where $1-P(T_{\max})\leq10^{-10}$.
The code is available online as a Python package, see \cite{maier_nwdiff_2017}.
For solving the integral Eq.~(\ref{eq:T_heur_ex}) we computed the
GMFPT of each node via Eq.~(\ref{eq:GMFPT_Laplacian}) with eigenvalues
and -vectors computed using the NumPy implementation \cite{jones_scipy:_2001}
of the standard algorithm for eigenvalue and -vector computation of
Hermitian matrices \cite{strang_linear_1980}.

\bibliographystyle{ieeetr}
\bibliography{cover_time}

\end{document}